\documentclass{elsarticle}
\usepackage[utf8]{inputenc}
\usepackage{amsmath}
\usepackage{amsfonts}
\usepackage{amssymb}
\usepackage{graphicx}
\usepackage{graphics}
\usepackage{textcomp}
\usepackage{booktabs}
\usepackage{paralist}
\usepackage{comment}

\usepackage{times}
\usepackage[T1]{fontenc}

\usepackage[numbers]{natbib}

\usepackage[usenames,dvipsnames]{xcolor}

\usepackage{soul}

\usepackage{listings}

\lstdefinelanguage{JavaScript}{
	basicstyle=\scriptsize\ttfamily,
	identifierstyle=\color{Black},
	sensitive=false,
	frame=lines,
	numbers=left,
	numberstyle=\tiny\color{Gray},
	numbersep=2pt,
	tabsize=2,
	showstringspaces=false,
	keywords={for, typeof, new, true, false, catch, return, null, catch, switch, var, if, in, while, else, case, break, of},
	keywordstyle=\color{Cerulean}\bfseries,
	ndkeywords={class, export, boolean, throw, implements, this},
	ndkeywordstyle=\color{Gray}\bfseries,
	stringstyle=\color{Red}\ttfamily,
	morestring=[b]',
	morestring=[b]",
	comment=[l]{//},
	morecomment=[s]{/*}{*/},
	commentstyle=\color{ForestGreen}\ttfamily
}

\lstdefinestyle{LARA}{
	language=JavaScript,
	keywords=[3]{import, aspectdef, select, end, apply, condition, input, output, dynamic},
	keywords=[4]{def, insert, exec, before, after,call},
	keywords=[5]{decl, func, fCall, file},
	keywordstyle={[3]\color{NavyBlue}\bfseries},
	keywordstyle={[4]\color{BurntOrange}\bfseries},
	keywordstyle={[5]\color{darkgray}\bfseries},
		literate={0}{{\textcolor{Red}{0}}}{1}%
		             {1}{{\textcolor{Red}{1}}}{1}%
		             {2}{{\textcolor{Red}{2}}}{1}%
		             {3}{{\textcolor{Red}{3}}}{1}%
		             {4}{{\textcolor{Red}{4}}}{1}%
		             {5}{{\textcolor{Red}{5}}}{1}%
		             {6}{{\textcolor{Red}{6}}}{1}%
		             {7}{{\textcolor{Red}{7}}}{1}%
		             {8}{{\textcolor{Red}{8}}}{1}%
		             {9}{{\textcolor{Red}{9}}}{1}%
		             {.0}{{\textcolor{Red}{.0}}}{2}
		             {.1}{{\textcolor{Red}{.1}}}{2}
		             {.2}{{\textcolor{Red}{.2}}}{2}%
		             {.3}{{\textcolor{Red}{.3}}}{2}%
		             {.4}{{\textcolor{Red}{.4}}}{2}%
		             {.5}{{\textcolor{Red}{.5}}}{2}%
		             {.6}{{\textcolor{Red}{.6}}}{2}%
		             {.7}{{\textcolor{Red}{.7}}}{2}%
		             {.8}{{\textcolor{Red}{.8}}}{2}%
		             {.9}{{\textcolor{Red}{.9}}}{2}%
}

\lstdefinestyle{MyC}{
	language=C,
	basicstyle=\scriptsize\ttfamily,
	identifierstyle=\color{Black},
	sensitive=false,
	frame=lines,
	numbers=left,
	numberstyle=\tiny\color{Gray},
	numbersep=2pt,
	tabsize=2,
	showstringspaces=false,
	stringstyle=\color{Red}\ttfamily,
	morestring=[b]',
	morestring=[b]",
	comment=[l]{//},
	morecomment=[s]{/*}{*/},
	commentstyle=\color{ForestGreen}\ttfamily,
	keywordstyle=\color{NavyBlue}\bfseries,
	literate={0}{{\textcolor{Red}{0}}}{1}%
	             {1}{{\textcolor{Red}{1}}}{1}%
	             {2}{{\textcolor{Red}{2}}}{1}%
	             {3}{{\textcolor{Red}{3}}}{1}%
	             {4}{{\textcolor{Red}{4}}}{1}%
	             {5}{{\textcolor{Red}{5}}}{1}%
	             {6}{{\textcolor{Red}{6}}}{1}%
	             {7}{{\textcolor{Red}{7}}}{1}%
	             {8}{{\textcolor{Red}{8}}}{1}%
	             {9}{{\textcolor{Red}{9}}}{1}%
	             {.0}{{\textcolor{Red}{.0}}}{2}
	             {.1}{{\textcolor{Red}{.1}}}{2}
	             {.2}{{\textcolor{Red}{.2}}}{2}%
	             {.3}{{\textcolor{Red}{.3}}}{2}%
	             {.4}{{\textcolor{Red}{.4}}}{2}%
	             {.5}{{\textcolor{Red}{.5}}}{2}%
	             {.6}{{\textcolor{Red}{.6}}}{2}%
	             {.7}{{\textcolor{Red}{.7}}}{2}%
	             {.8}{{\textcolor{Red}{.8}}}{2}%
	             {.9}{{\textcolor{Red}{.9}}}{2}%
}

\usepackage{xspace}
\usepackage{todonotes}

\usepackage[bookmarks, colorlinks, breaklinks]{hyperref}  
\hypersetup{linkcolor=blue,citecolor=blue,filecolor=black,urlcolor=MidnightBlue} 

\usepackage{cleveref}

\newcommand{\libVersioningCompiler}{\textsc{libVersioningCompiler}\xspace}
\newcommand{\libVC}{\textsc{libVC}\xspace}

\title{The ANTAREX Domain Specific Language for High Performance Computing}

\author[polimi]{Cristina Silvano}
\ead[polimi]{name.surname@polimi.it}
\author[polimi]{Giovanni Agosta}
\author[unibo]{Andrea Bartolini}
\author[dompe]{Andrea R. Beccari}
\author[ethz]{Luca Benini}
\author[irisa]{Lo\"{\i}c Besnard}
\author[uporto]{Jo\~ao Bispo}
\author[sygic]{Radim Cmar}
\author[uporto]{Jo\~ao M. P. Cardoso}
\author[cineca]{Carlo Cavazzoni}
\author[unibo]{Daniele Cesarini}
\author[polimi]{Stefano Cherubin}
\author[cineca]{Federico Ficarelli}
\author[polimi]{Davide Gadioli}
\author[it4i]{Martin Golasowski}
\author[ethz]{Antonio Libri}
\author[it4i]{Jan Martinovi\v{c}}
\author[polimi]{Gianluca Palermo}
\author[uporto]{Pedro Pinto}
\author[inria]{Erven Rohou}
\author[it4i]{Kate\v{r}ina Slaninov\'a}
\author[polimi]{Emanuele Vitali}

\address[polimi]{DEIB -- Politecnico di Milano, Piazza Leonardo da Vinci 32, Milano, Italy} 

\address[ethz]{IIS -- Eidgen\"ossische Technische Hochschule Z\"urich}

\address[unibo]{Alma Mater Studiorum - University of Bologna}

\address[dompe]{Domp\'e Farmaceutici SpA}

\address[uporto]{FEUP -- Universidade do Porto}

\address[cineca]{CINECA}

\address[it4i]{IT4Innovations, VSB -- Technical University of Ostrava}

\address[inria]{INRIA Rennes}

\address[sygic]{Sygic}

\address[irisa]{IRISA/CNRS}

\begin{document}

\begin{abstract}
The ANTAREX project relies on a Domain Specific Language (DSL) based on Aspect Oriented Programming (AOP) concepts to allow applications to enforce extra functional properties such as energy-efficiency and performance and to optimize Quality of Service (QoS) in an adaptive way.
The DSL approach allows the definition of energy-efficiency, performance, and adaptivity strategies as well as their enforcement at runtime through application autotuning and resource and power management.
In this paper, we present an overview of the key outcome of the project, the ANTAREX DSL, and some of its capabilities through a number of examples, including how the DSL is applied in the context of the project use cases.
\end{abstract}


\maketitle

\section{Introduction}
\label{sec:intro}
High Performance Computing (HPC) is a strategic asset vigorously pursued by both state actors, supra-national entities, and major industrial players in fields such as finance or oil \& gas~\cite{curley14hpc}.
A veritable arms race has been waged for decades to build the fastest HPC machine, first in terms of pure floating pooint operations per second (FLOPS), then, with the growing difficulty of supplying power to large HPC centers, in terms of Flops/watt.
Currently, the goal is to reach Exascale level ($10^{18}$ FLOPS) within the $2023-24$ timeframe -- with a $\times$$1000$ improvement over Petascale, reached in $2009$.
The current Top500\footnote{www.top500.org, November 2018} and Green500\footnote{www.green500.org, November 2018} lists are dominated respectively by IBM's Summit machine, installed at the Oak Ridge National Laboratory (USA), with over 200 PetaFlops, and Japan's Shoubu system B, with a power efficiency of 17.6 GFlops/watt\footnote{Summit is also ranked 3rd in the Green500, whereas Shoubu is a comparatively smaller machine}.
To reach these levels of performance and power efficiency, heterogeneous computing architectures are critical, in particular through GPGPUs.
Traditional NVIDIA GPGPUs are featured heavily in both lists, though Shoubu's ZettaScaler architecture uses PEZY's PEZY-SC2 instead.
The dominance of heterogeneous systems in the Green500 list is expected to continue for the next coming years to reach the target of 20MW Exascale supercomputers set by the DARPA. 

When considering the design, development, and deplyoment of HPC applications, each increase in HPC machine size and heterogeneity imposes an increased burden on the developer. 
Nowadays, few developers have both the domain knowledge and HPC expertise to produce a fully optimised application, and a na\"ively parallelised application may easily incur in parallelisation bottlenecks and scalability issues.
To address this issue, the current development model assumes that the domain experts do not attempt to optimise their applications on their own, but rely on help from HPC experts, typically part of the HPC center staff, to turn their baseline application into an HPC-enabled one.
This task requires to handle parallelisation, offloading to heterogeneous resources, as well as performance analysis and tuning.
For the latter task, it is worth noting that for large scale applications it is oftern difficult to make predictions on performance based on small-scale runs, so runtime autotuning is desirable.

While many approaches to programming languages and models for HPC attempt to simplify the development process by reducing or removing the need for the HPC expert, the recently completed ANTAREX project~\cite{silvano2017antarex,silvano2018antarex:dsd} fully embraces this split development process, and aims at supporting it.
To this end, we introduce the ANTAREX DSL, a Domain Specific Language (DSL) to express the application self-adaptivity and to runtime manage and autotune applications for green heterogeneous HPC systems up to the Exascale level.
The ANTAREX DSL allows the introduction of a separation of concerns, where self-adaptivity and energy efficient strategies, managed by the HPC expert, are specified separately from the application functionalities, developed instead by the domain expert. 
The ANTAREX DSL is based on Aspect Oriented Programming (AOP) principles, embodied in the LARA language~\cite{cardoso_lara:_2012, cardoso_performance-driven_2014}, which effectively support the separation of concern without burdening the domain expert with the need to learn a new language (as they simply develop in C/C++).
Since HPC experts already use a variety of scripts and tools to support their work, the ANTAREX DSL also provides a way to unify them under a single interface.

To demonstrate the effectiveness of the ANTAREX DSL and its associated tools, the project employs two use cases taken from highly relevant HPC application scenarios: 
\begin{enumerate}
\item a biopharmaceutical application for drug discovery developed by one of the leading European companies in the field, Domp\'e, and deployed on the 1.21 PetaFlops heterogeneous NeXtScale Intel-based IBM system at CINECA; 
\item a self-adaptive navigation system for smart cities developed by the top European navigation software company, Sygic, and deployed on the server-side on the 1.46 PetaFlops heterogeneous  Intel\textregistered\  Xeon Phi\texttrademark\ based system provided by IT4Innovations National Supercomputing Center. 
\end{enumerate}

These scenario also reproduce the aforementioned development model, as each HPC user company is supported by the staff of the HPC center where the application is deployed.


The remaining partners of the ANTAREX Consortium comprise a wealth of expertise in DSL development, code optimisation, energy efficiency, and autotuning. Four top-ranked academic and research partners (Politecnico di Milano, ETHZ Zurich, University of Porto and INRIA) are complemented by the Italian Tier-0 Supercomputing Center (CINECA), the Tier-1 Czech National Supercomputing Center (IT4Innovations) and two industrial application providers, one of the leading biopharmaceutical companies in Europe (Domp\'e) and the top European navigation software company (Sygic). 
Politecnico di Milano, the largest Technical University in Italy, played the role of Project Coordinator.

\paragraph{The ANTAREX Approach}
\begin{figure}[ht]
\centering
\resizebox{\columnwidth}{!}{\includegraphics{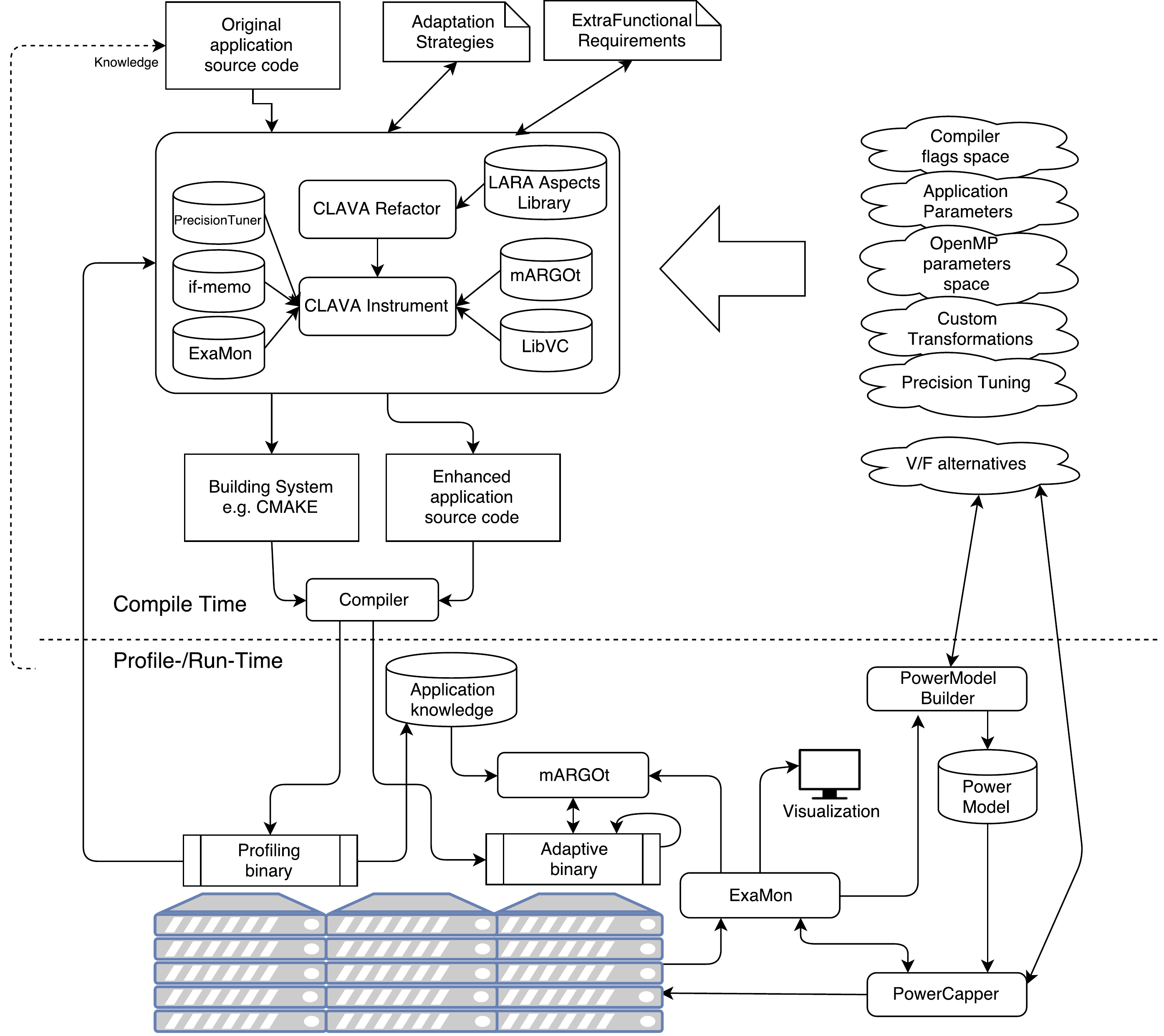}}
\caption{\label{fig:toolflow}The ANTAREX Tool Flow, comprised of a compile-time part, where the instrumentation of the functional code is performed based on the strategies expressed in the ANTAREX DSL, and a run-time part, where the generated adaptive binary is monitored during its run on the system and optimised by tuning its parameters, including selection of different code versions.}
\end{figure}

The ANTAREX approach and related tool flow, as shown in Figure~\ref{fig:toolflow}, operate both at design-time and runtime. 
The application functionality is expressed through C/C++ code (possibly including legacy code), whereas the extra-functional aspects of the application, including parallelisation, mapping, and adaptivity strategies, are expressed through DSL code (based on LARA) developed in the project. 
As a result, the expression of such aspects is fully decoupled from the functional code. 
The \emph{Clava} tool is the centerpoint of the compile-time phase, performing a refactoring of the application code based on the LARA aspects, and instrumenting it with the necessary calls to other components of the tool flow.

The ANTAREX compilation flow leverages a runtime phase with compilation steps, through the use of partial dynamic compilation techniques enabled by \emph{libVC}. 
The application autotuning, performed via the \emph{mARGOt} tool, is delayed to the runtime phase, where the software knobs (application parameters, code transformations and code variants) are configured according to the runtime information coming from application self-monitoring as well as from system monitoring performed by the \emph{ExaMon} tool.
Finally, the runtime power manager, \emph{PowerCapper}, is used to control the resource usage for the underlying computing infrastructure given the changing conditions.
At runtime, the application control code, thanks to the design-time phase, now contains also runtime monitoring and adaptivity strategy code derived from the DSL extra-functional specification. 
Thus, the application is continuously monitored to guarantee the required Service Level Agreement (SLA), while communication with the runtime resource-manager takes place to control the amount of processing resources needed by the application. 
The application monitoring and autotuning is supported by a runtime layer implementing an application level collect-analyse-decide-act loop.

\paragraph{Organization of the paper}
The rest of this paper is organized as follows.
In Section~\ref{sec:tech} we review the technology portfolio provided by the ANTAREX tool flow.
In Section~\ref{sec:eval} we provide an assessment of the impact of the proposed DSL on application specifications, while Sections~\ref{sec:appl1} and~\ref{sec:appl} show of how the Tool Flow has been applied the two use cases of the project.
Finally, in Section~\ref{sec:relw} we compare the ANTAREX DSL with related works from the recent literature, and in Section~\ref{sec:conc} we draw some conclusions.



\section{ANTAREX Technology Portfolio}
\label{sec:tech}
In this Section, we first provide an overview of the ANTAREX DSL itself, and then we show how the DSL is used to apply each of the main components of the Tool Flow, providing examples of the code needed in each case to allow the reader to understand the level of complexity involved.

\subsection{The ANTAREX DSL}
\label{sec:dsl}
%
HPC applications might profit from adapting to operational and situational conditions, such as changes in contextual information (e.g., workloads), in requirements (e.g., deadlines, energy), and in availability of resources (e.g., connectivity, number of processor nodes available). 
A simplistic approach to both adaptation specification and implementation (see, e.g., \cite{Floch:2006}) employs hard coding of, e.g., conditional expressions and parameterizations. In our approach, the specification of runtime adaptability strategies relies on a DSL implementing key concepts from Aspect-Oriented Programming (AOP)~\cite{irwin_aspect-oriented_1997}, mainly specifying adaptation concerns, targeting specific execution points, separately from the primary functionality of the application, with minimum or no changes to the application source code. 


Our approach is based on the idea that certain application/system requirements (e.g., target-dependent optimizations, adaptivity behavior and concerns) should be specified separately from the source code that defines the main functionality. Those requirements are expressed as DSL aspects that embody strategies. An extra compilation step, performed by a \textit{weaver}, merges the original source code and aspects into the intended program~\cite{elrad_aspect-oriented_2001}. Using aspects to separate concerns from the core objective of the program can result in cleaner programs and increased productivity (e.g., reusability of strategies).
As the development process of HPC applications typically involves two types of experts (application-domain experts and HPC system architects) that split their responsibilities along the boundary of functional description and extra-functional aspects, our DSL-aided toolflow provides a suitable approach for helping to express their concerns.

The ANTAREX DSL relies on the already existing DSL technology LARA~\cite{cardoso_lara:_2012, cardoso_performance-driven_2014}.
In particular, the LARA technology provides a framework that we adopted to implement the ANTAREX aspects and APIs.
Moreover, we developed other LARA-related tools such as the \emph{Clava}\footnote{\href{https://github.com/specs-feup/clava}{https://github.com/specs-feup/clava}} weaver to leverage the rest of the ANTAREX tool flow.

LARA is a programming language that allows developers to capture non-functional requirements and concerns in the form of strategies, which are decoupled from the functional description of the application. Compared to other approaches that usually focus on code injection (e.g., \cite{spinczyk_aspectc++:_2002}), LARA provides access to other types of actions, e.g., code refactoring, compiler optimizations, and inclusion of additional information, all of which can guide compilers to generate more efficient implementations.
%
Additional types of actions may be defined in the language specification and associated weaver, such as software/hardware partitioning~\cite{cardoso_controlling_2013} or compiler optimization sequences~\cite{nobre_use_2015}.
%
One important feature of the LARA-aided source-to-source compiler developed in ANTAREX is the capability to refactor the code of the application in order to expose adaptivity behavior and/or adaptivity design points that can be explored by the ANTAREX autotuning component.
In the following sections we show illustrative examples\footnote{Complete working versions for all examples can be found in \href{https://github.com/specs-feup/specs-lara/tree/master/2018 DSD}{https://github.com/specs-feup/specs-lara/tree/master/2018\%20DSD}} of some of the strategies that can be specified using LARA in the context of a source-to-source compiler and currently used for one of the use cases.

\subsection{Precision Tuning}
\label{sec:fixedp}

Error-tolerating applications are increasingly common in the emerging field of real-time HPC, allowing to trade-off precision for performance and/or energy. 
Thus, recent works investigated the use of customized precision in HPC as a way to provide a breakthrough in power and performance. 
We developed a set of LARA aspects enabling mixed precision tuning on C/C++ and OpenCL kernels.
In our precision tuning we combine an adaptive selection of floating and fixed point arithmetic, targeting HPC applications.

Figure~\ref{fig:precision} presents part of a LARA strategy that changes all declarations of a certain type to a target type (e.g., from double to float) for a given function. We note, however, that a practical and reusable aspect needs to deal with further issues, such as the cloning of functions whose types we want to change but are also called by other unrelated functions in the code, assignments of constants, casts, recursion, changing functions definitions and library functions to the ones related to the type used (e.g., \texttt{sqrtf} vs \texttt{sqrt} in Math.h), etc. In this example, \texttt{changeType} is a function that analyzes and changes compound types, such as \texttt{double*} and \texttt{double[]}. If the type described in \texttt{\$old} is found inside the type of the declaration, it is replaced with the type described in \texttt{\$new}. To be more specific, if \texttt{\$old} is \texttt{double}, \texttt{\$new} is \texttt{float} and \texttt{\$decl.type} is \texttt{double*}, the type of the declaration will be changed to \texttt{float*}. If the original declaration type does not contain the \texttt{\$old} type, it is not changed.

\begin{figure}[h]
	\begin{center}
\begin{lstlisting}[boxpos=b,style=LARA]
aspectdef ChangePrecision

	input $func, $old, $new end
	
	/* change type of variable declarations found
	 *  inside the function and parameters */
	select $func.decl end
	apply
		var changedType = changeType($decl.type, $old, $new);
		def type = changedType;
	end

	/* do the same with the function return type ... */
	var $returnType = $func.functionType.returnType;
	$func.functionType.def returnType =
				changeType($returnType, $old, $new);
end
\end{lstlisting}
	\end{center}
	\caption{Example of LARA aspect to change the types of variables declared inside a given function.}
	\label{fig:precision}
\end{figure}

A LARA aspect consists of three main steps. Firstly, one captures the points of interest in the code using a \texttt{select} statement, which in this example selects variable declarations. Then, using the \texttt{apply} statement, one acts over the selected program points. In this case, it will define the types of the captured declared variables, using the \texttt{type} attribute. Finally, we can then specify conditions to constrain the execution of the \texttt{apply} (i.e., only if the declared variable has a specific type). This can be done via conditional statements (\texttt{if}s) as well as via special \texttt{condition} blocks that constrain the entire \texttt{apply}. LARA promotes modularity and aspect reuse, and supports embedding JavaScript code, to specify more sophisticated strategies.
As shown in~\cite{Nobre2018PARMADITAM}, we support exploration of mixed precision OpenCL kernels by using half, single, and double precision floating point data types.
We additionally support fixed point representations through a custom C++ template-based implementation for HPC systems, which has already been used in~\cite{cherubin2017ParCo}.
In both cases the LARA aspects automatically insert code for proper type conversion before and after the critical section that has been converted to exploit a reduced precision data type.

The LARA aspect in Figure~\ref{fig:prec_tuning} shows the generation of different mixed-precision versions to be dynamically evaluated.
It is possible to specify -- as input of the aspect -- the number of mix combination to generate, and a rule set to filter out precision mix combinations which are very likely to lead to useless and/or not efficient results.
We exploit programmer's application domain knowledge by relying on them to define test cases to evaluate the different code versions at runtime.
LARA automatically inserts code to dynamically perform the exploration over the space of the generated versions with different precision mix.

\begin{figure}[ht]
	\begin{center}
\begin{lstlisting}[boxpos=b,style=LARA]
aspectdef HalfPrecisionOpenCL
	input 
		combinationFilter = [],
		maxVersions = undefined
	end
	
	// List of float and double vars in the OpenCL kernel
	call result : OpenCLVariablesToTest;
	var variablesToTest = result.variablesToTest;
	// Sequence generator
	var sequenceGenerator = new SequentialCombinations(
		variablesToTest.length, maxVersions);
	var counter = 0;
	while(sequenceGenerator.hasNext()) {
		Clava.pushAst(); // Save current AST
		
		// Get a new combination of variables
		var combination = sequenceGenerator.next();
		var lastSeed = sequenceGenerator.getLastSeed();
		if(!isCombinationValid(lastSeed, combinationFilter))
			continue;	
		// Change the builtin type of the variables
		for(var index of combination) {
			var $vardecl = Clava.findJp(variablesToTest[index]);
			changeTypeToHalf($vardecl);
		}
		call addHalfPragma(); // Enable half-precision
		var outputFolder = createFolder(lastSeed, 
			variablesToTest.length, counter);	
		Clava.writeCode(outputFolder); // Generate code
		Clava.popAst(); // Restore previous AST tree
		counter++; // Increase counter
	}
end
\end{lstlisting}
	\end{center}
	\caption{Example of LARA aspect that generates different precision mix versions of the same OpenCL kernel.}
	\label{fig:prec_tuning}
\end{figure}

\subsection{Code Versioning}
\label{sec:libvc}

One of the strategies supported in the ANTAREX toolflow is the capability to generate versions of a function and to select the one that satisfies certain requirements at runtime.
Figure~\ref{fig:clone} shows an aspect that clones a set of functions and changes the types of the newly generated clones. Each clone has the same name as the original with the addition of a provided suffix. We start with a single user-defined function which is cloned by the aspect \texttt{CloneFunction} (called in line 13). Then, it recursively traverses calls to other functions inside the clone and generates a clone for each of them. Inside the clones, calls to the original functions are changed to calls to the clones instead, building a new call tree with the generated clones. At the end of the aspect \texttt{CreateFloatVersion} (lines 16--17,) we use the previously defined \texttt{ChangePrecision} aspect to change the types of all generate clones.

\begin{figure}[h]
	\begin{center}
\begin{lstlisting}[boxpos=b,style=LARA]
import ChangePrecision;
import clava.ClavaJoinPoints;

aspectdef CreateFloatVersion
	input $func, suffix end
	output $clonedFunc end
	
	$double = ClavaJoinPoints.builtinType('double');
	$float = ClavaJoinPoints.builtinType('float');
	
	/* clone the target functions and the child calls */
	var clonedFuncs = {};
	call cloned : CloneFunction($func, suffix, clonedFuncs);

	/* change the precision of the cloned function */
	for($clonedFunc of clonedFuncs)
		call ChangePrecision($clonedFunc, $double, $float);
	
	$clonedFunc = cloned.$clonedFunc;	
end
\end{lstlisting}
	\end{center}
	\caption{\label{fig:clone}Example of LARA aspect to clone an existing function and change the type of the clone.}
\end{figure}

The aspect \texttt{Multiversion} -- in Figure~\ref{fig:main} -- adapts the source code of the application in order to call the original version of a function or a generated cloned version with a different type, according to the value of a parameter given by the autotuner at runtime. The main aspect calls the previously shown aspect, \texttt{CreateFloatVersion}, which clones the target function and every other function it uses, while also changing their variable types from \texttt{double} to \texttt{float} (using the aspects presented in Figure~\ref{fig:clone} and Figure~\ref{fig:precision}). This is performed in lines 8--9 of the example.
From lines 13 to 34, the \texttt{Multiversion} aspect generates and inserts code in the application that is used as switching mechanism between the two versions. It starts by declaring a variable to be used as a knob by the autotuner, then it generates the code for a switch statement and replaces the statement containing the original call with the generated switch code. Finally, in lines 36--38, the aspect surrounds both calls (original and float version) with timing code. An excerpt of the resulting \textsc{C} code can bee seen in Figure~\ref{fig:cresult}.

\begin{figure}[h]
	\begin{center}
\begin{lstlisting}[boxpos=b,style=LARA]
import CreateFloatVersion;
import lara.code.Timer;
import clava.ClavaJoinPoints;

aspectdef Multiversion
	input $func, knobName end
	
	call fVersion : CreateFloatVersion($func, "_f");
	var $floatFunc = fVersion.$clonedFunc;
	var timer = new Timer();
	
	 /* Identify call by name... */
	 select function.body.stmt.call{$func.name} end
	 apply
		/* ... and by type signature */
		if(!$func.functionType.equals($call.functionType))
			continue;

		/* Add knob for choosing the version */
		$int = ClavaJoinPoints.builtinType('int');
		$body.exec addLocal(knobName, $int, 0);

		/* create float declaration for first argument */
		var $arg = createFloatArg($call.args[0]);
		/* Create call based on float version of function */
		$floatFunc.exec $fCall : newCall([$arg, $call.args[1]]);
		/* Copy current call */
		$call.exec $callCopy : copy();
			
		/* Create switch */
		var $condition = ClavaJoinPoints.exprLiteral(knobName);
		var switchCases = {0: $callCopy, 1: $fCall};
		call switchJp : CreateSwitch($condition, switchCases);				
		$stmt.exec replaceWith(switchJp.$switch);

		/* Time calls to both original and float functions*/
		timer.time($callCopy, "Original time:");
		timer.time($fCall, "Float time:");
	 end
end
\end{lstlisting}
	\end{center}
	\caption{Example of LARA aspect that generates an alternative version of a function and inserts a mechanism in the code to switch between versions.}
	\label{fig:main}
\end{figure}

\begin{figure}[h]
	\begin{center}
\begin{lstlisting}[boxpos=b,style=MyC]
switch (version) {
	case 0: {
		clock_gettime(CLOCK_MONOTONIC, &time_start_0);
		SumOfInternalDistances(atoms, 1000);
		clock_gettime(CLOCK_MONOTONIC, &time_end_0);
		double time_0 = calc_time(time_start_0, time_end_0); 
		printf("Original time:%fms\n", time_0);
	}
	break;
	case 1: {
		clock_gettime(CLOCK_MONOTONIC, &time_start_1);
		SumOfInternalDistances_f(atoms_f, 1000);
		clock_gettime(CLOCK_MONOTONIC, &time_end_1);
		double time_1 = calc_time(time_start_1, time_end_1); 
		printf("Float time::%fms\n", time_1);
	}
	break;
}
\end{lstlisting}
	\end{center}
	\caption{Excerpt of the \textsc{C} code resulting from the generation of alternative code versions.}
	\label{fig:cresult}
\end{figure}

In the ANTAREX toolflow, the capability of providing several versions of the same function is not limited to static features.
\libVersioningCompiler~\cite{cherubin2018libVC,festa2018continuos} (abbreviated \libVC) is an open-source C++ library designed to support the dynamic generation and versioning of multiple versions of the same compute kernel in a HPC scenario.
It can be used to support continuous optimization, code specialization based on the input data or on workload changes, or to dynamically adjust the application, without the burden of a full just-in-time compiler.
\libVC allows a C/C++ compute kernel to be dynamically compiled multiple times while the program is running, so that different specialized versions of the code can be generated and invoked.
Each specialized version can be versioned for later reuse.
When the optimal parametrization of the compiler depends on the program workload, the ability to switch at runtime between different versions of the same code can provide significant benefits~\cite{chen2010evaluating,tartara2013continuous}.
While such versions can be generated statically in the general case, in HPC execution times can be so long that exhaustive profiling may not be feasible.
\libVC instead enables the exploration and tuning of the parameter space of the compiler at runtime.


Figure~\ref{fig:libVC} shows an example of usage of \libVC through LARA, which demonstrates how to specialize a function. The user provides this aspect with a target function call and a set of compilation options. These include compiler flags and possible compiler definitions, e.g., data discovered at runtime, which is used as a compile-time constant in the new version. Based on the target function call, the aspect finds the function definition which is passed to the library. After the options are set, the original function call is replaced with a call of the newly compiled and loaded specialized version of the kernel.

\begin{figure}[h]
	\begin{center}
\begin{lstlisting}[boxpos=b,style=LARA]
import antarex.libvc.LibVC;

aspectdef SimpleLibVC

	input
		name, $target, options
	end

	var $function = $target.definition;
	var lvc = new LibVC($function, {logFile:"log.txt"}, name);

	var lvcOptions = new LibVCOptions();
	for (var o of options) {
		lvcOptions.addOptionLiteral(o.name, o.value, o.value);
	}
	lvc.setOptions(lvcOptions);

	lvc.setErrorStrategyExit();

	lvc.replaceCall($target);
end
\end{lstlisting}
	\end{center}
	\caption{Example of LARA aspect to replace a function call to a kernel with a call to a dynamically generated version of that kernel.}
	\label{fig:libVC}
\end{figure}

It is worth noting that the combination of LARA and \libVC can also be used to support compiler flag selection and phase-ordering both statically and dynamically~\cite{ashouri2016cobayn,ashouri2017micomp}.

\subsection{Memoization}
\label{sec:memoi}
Memoization is a technique that trades off computation time for memory space by storing computed values and reusing them instead of recomputing them.
It has been shown to be effective in improving performance and reducing energy consumption in large scale applications~\cite{agosta2011dynamic}.
We introduce in this section a memoization technique integrated in the ANTAREX toolflow.
Performance can be improved by caching results of pure functions (i.e. deterministic functions without side effects), 
and retrieving them instead of recomputing a result. We have implemented the work of~\cite{suresh:hal-01423811} generalized for C++ and aided with extensions regarding user/developer flexibility.
 We describe here only the principles of this technique and more details can be found in~\cite{suresh:hal-01178085}~\cite{suresh:hal-01423811}.

\begin{figure}[h]
  \begin{center}
     \begin{lstlisting}[boxpos=b,style=myC]
        float foo (float p) {
        /* code of foo without side effects */
        }

        float foo_wrapper(float p)
            {
              float r;
              /* already in the table ? */
              if (lookup_table(p, &r)) return r;
              /* calling the original function */
              r = foo(p); 
              /* updating the table or not */
              update_table(p, r); 
              return r;

     \end{lstlisting}
  \end{center}
  \caption{A memoizable C function and its wrapper.}
  \label{fig:fmemoi}
\end{figure}

Consider a memoizable C function \verb?foo? as shown in Figure~\ref{fig:fmemoi}. The memoization consists in:
\begin{enumerate}  
\item the insertion of a wrapper function \verb?foo_wrapper? and an associated table.
\item The substitution of the references to \verb?foo? by \verb?foo_wrapper? in the application.
\end{enumerate}

This technique has been extended for C++ memoizable methods and takes into account the mangling, the overloading, and the references to the objects.
Memoization is proposed in the ANTAREX project by relying on  aspects programmed using the DSL.
The advantage of these aspects is that the memoization is integrated into the application without requiring user modifications of the source code.
The code generated by Clava is then compiled and linked with the associated generated memoization library.

\begin{figure}[h]
  \begin{center}
\begin{lstlisting}[boxpos=b,style=LARA]
aspectdef Memoize_Method_overloading_ARGS
input 
  aClass,     // Name of a class
  aMethod,    // Name of a method of the class aClass
  pType,      // Name of the selected type
  nbArgs,     // Number of parameters of the method
  fileToLoad, // filename for init of the table, or 'none'
  FullOffLine,// yes for a fully offline strategy
  FileToSave, // filemane to save the table, or 'none'
  Replace,    // Always replace in case of collisions
  approx,     // Number of bits to delete for approximation.
  tsize       // Size of the internal table.
end
// Control on the parameters of the aspect: nbArgs in [1,3]
   ...
// Searching the method.
var MethodToMemoize, found=false; 
select class{aClass}.method{aMethod} end  
apply
 if (! found) {
   found = isTheSelectedMethod($method, nbArgs, pType);
   if (found) MethodToMemoize=$method;
 }
end
if (!found) 
 { /* message to the user */}
else {
 GenCode_CPP_Memoization(aClass, aMethod, pType, nbArgs, 
 fileToLoad, FullOffLine, FileToSave, Replace, approx, tsize);
 call CPP_UpdateCallMemoization(aClass, aMethod, pType, nbArgs);
}
end	
\end{lstlisting}
  \end{center}
  \caption{An example of LARA aspect defined for the memoization.}
  \label{fig:memoiaspect}
\end{figure}
An example of a LARA aspect for memoization is shown in Figure~\ref{fig:memoiaspect}. It defines the memoization (lines 1-13) of a method (\verb?aMethod?) 
of a class (\verb?aClass?) with \verb?nbArg? parameters of same type as the returned type (\verb?Type?).
Note that the inputs \verb?nbArg? and  \verb?Type? are required to manage the overloading of the object-oriented languages such that C++.
Other parameters (from line~4) are provided to improve several memoization approaches. 
For examples, the user can specify (1) the policy in case of conflicts regarding the same table entry (line~11): replacement or not in case of conflict to the same entry of the table for different parameters of the memoized function, and (2) the size of the table (line~15).
After some verifications, not detailed here, on the parameters (lines 14-15), the method is searched (lines 17-24). 
Then, in case of success, the code of the wrapper is added (line 28) to produce the memoization library, and (line 30) the code of the application is modified 
for calling the created ``wrapper'', this wrapper is also declared as a new method of the class.

Moreover, some variables are exposed for autotuning in the memoization library. For each function or method, a variable that manages the 
dynamical "stop/run" of the memoization is exposed, as well as the variable that manages the policy to use in case of conflict to the table.
To be complete about the memoization, a LARA aspect is proposed to automatically detect the memoizable functions or methods.
Then the user may decide to apply or not the memoization on these selected elements.

\subsection{Self-Adaptivity \& Autotuning}
\label{sec:autotuning}
In ANTAREX, we consider each application's function as a parametric function that elaborates input data to produce an output (i.e., $o = f(i,k_1,\ldots,k_n)$ ), with associated extra-functional requirements. In this context, the parameters of the function ($k_1,\ldots,k_n$) are software-knobs that modify the behavior of the application (e.g., parallelism level or the number of trials in a MonteCarlo simulation).
The main goal of mARGOt~\footnote{\href{https://gitlab.com/margot_project/core}{https://gitlab.com/margot\_project/core}}~\cite{gadioli2015application} is to enhance an application with an adaptive layer, aiming at tuning the software knobs to satisfy the application requirements at runtime.
To achieve this goal, the mARGOt dynamic autotuning framework developed in ANTAREX is based on the MAPE-K feedback loop~\cite{MAPEK}.
In particular, it relies on an application knowledge, derived either at deploy time or at runtime, that states the expected behavior of the extra-functional properties of interest.
To adapt, on one hand mARGOt uses runtime observations as feedback information for reacting to the evolution of the execution context.
On the other hand, it considers features of the actual input to adapt in a more proactive fashion.
Moreover, the framework is designed to be flexible, defining the application requirements as a multi-objective constrained optimization problem that might change at runtime.

To hide the complexity of the application enhancement, we use LARA aspects for configuring mARGOt and for instrumenting the code with related API. 
Figure~\ref{fig:laraMARGOT} provides a simple example of a LARA aspect where mARGOt has been configured (lines 5-20) to actuate on a software knob \emph{Knob1} and target \emph{error} and \emph{throughput} metrics~\cite{SOCRATES}. 
In particular, the optimization problem has been defined as the maximization of the \emph{throughput} while keeping the \emph{error} under a certain threshold.
The last part of the aspect (lines 23-27) is devoted to the actual code enhancement including the needed mARGOt call for initializing the framework and for updating the application configuration.
The declarative nature of the LARA library developed for integrating mARGOt simplifies its usage hiding all the details of the framework.

\begin{figure}[h]
	\begin{center}
\begin{lstlisting}[boxpos=b,style=LARA]
aspectdef mARGOt_Aspect
/* Input: TargetFunctionCall*/
input targetCallName end
/* mARGOt configuration */
  var config = new MargotConfig();
  var targetBlock = config.newBlock($targetCallName);
  targetBlock.addKnob('Knob1', 'knob1', 'int');
  targetBlock.addMetric('error', 'float');
  targetBlock.addMetric('throughput', 'float');
  targetBlock.addMetricGoal('my_error_goal', 
    MargotCFun.LE, 0.03, 'error');

  /* optimization problem */
  var problem = targetBlock.newState('defaultState');
  problem.maximizeMetric('throughput');
  problem.subjectTo('my_error_goal');
    
  /* generate the information needed 
     for enhancing the application code */
  codegen = MargotCodeGen.fromConfig(config, $targetCallName);
    
/* Target function call identification */
select stmt.call{targetName} end
   
/* Add mARGOt calls*/
codegen.init($call);
codegen.update($call);
end
\end{lstlisting}
	\end{center}
	\caption{Example of a LARA aspect for autotuner configuration and code enhancement.}
	\label{fig:laraMARGOT}
\end{figure}

\subsection{Monitoring}
\label{sec:examon}
Today processing elements embed the capability of monitoring their current performance efficiency by inspecting the utilization of the micro-architectural components as well as a set of physical parameters (i.e., power consumption, temperature, etc). These metrics are accessible through hardware performance counters which in x86 systems can be read by privilege users, thus creating practical problems for user-space libraries to access them. Moreover, in addition to sensors which can be read directly from the software running on the core itself, supercomputing machines embed sensors external to the computing elements but relevant to the overall energy-efficiency. These elements include the node and rack cooling components as well as environmental parameters such as the room and ambient temperature. 
In ANTAREX, we developed ExaMon\cite{EXAMON} (\emph{Exa}scale \emph{Mon}itoring) to virtualise the performance and power monitoring access in a distributed environment. ExaMon decouples the sensor readings from the sensor value usage. Indeed, ExaMon uses a scalable approach were each sensor is associated to a sensing agent which periodically collects the metrics and sends the measured value with a synchronized time-stamp to an external data broker.
The data broker organises the incoming data in communication channels with an associated topic. 
Every new message on a specific topic is then broadcast to the related subscribers, according to a list kept by the broker.
The subscriber registers a callback function to the given topic which is called every time a new message is received. 
To let LARA take advantage of this monitoring mechanism we have designed the Collector API, which allow the initialization of the Collector component associated with a specific topic that keeps an internal state of the remote sensor updated. 
This internal state can then be queried asynchronously by the Collector API to gather its value. 
LARA aspects have been designed to embed the Collector API and to make the application code self-aware.

Figure~\ref{fig:laraExamon} shows a usage example of ExaMon through LARA, which subscribes to a topic on a given broker and inserts a logging message in the application. To define the connection information, the user needs to provide the address to connect to, as well as the name of the topic to subscribe. As for the integration in the original application code the user needs to provide a target function, where the collector will be managed, and a target statement, where the query of the data and logging will be performed.

\begin{figure}[ht]
	\begin{center}
\begin{lstlisting}[boxpos=b,style=LARA]
import antarex.examon.Examon;
import lara.code.Logger;

aspectdef SimpleExamon

	input
		name, ip, topic, $manageFunction, $targetStmt
	end

	var broker = new ExamonBroker(ip);

	var exa = new ExamonCollector(name, topic);

	// manage the collector on the target function
	select $manageFunction.body end
	apply
		exa.init(broker, $body);
		exa.start($body);

		exa.end($body);
		exa.clean($body);
	end

	// get the value and use it in the target stmt
	exa.get($targetStmt);

	// get the last stmt of the scope of the target stmt
	var $lastStmt = $targetStmt.ancestor("scope").lastStmt;
	// Create printf for time and data
	var logger = new Logger();
	logger.ln().text("Time=").double(getTimeExpr(exa))
		.text("[s], data=").double(exa.getMean()).ln();
	// Add printf after last stmt	
	logger.log($lastStmt);
end
\end{lstlisting}
	\end{center}
	\caption{Example of a LARA aspect integrate an ExaMon collector into an application.}
	\label{fig:laraExamon}
\end{figure}

\subsection{Power Capping}
\label{sec:pcap}
Today's computing elements and nodes are power limited. 
For this reason, state-of-the-art processing elements embed the capability of fine-tuning their performance to control dynamically their power consumption. 
This includes dynamic scaling of voltage and frequency, and power gating for the main architectural blocks of the processing elements, but also some feedback control logic to keep the total power consumption of the processing element within a safe power budget. 
This logic in x86 systems is named RAPL~\cite{RAPL}.
Demanding the power control of the processing element entirely to RAPL may not be the best choice. 
Indeed, it has been recently discovered that RAPL is application agnostic and thus tends to waist power in application phases which exhibit IOs or memory slacks. 
Under these circumstances there are operating points that proved to be more energy efficient than the ones selected by RAPL while still respecting the same power budget~\cite{CESARINI_ANDARE}. 
However, these are only viable if the power capping logic is aware of the application requirements. 
To do so, we have developed a new power capping run-time based on a set of user space APIs which can be used to define a relative priority for the given task currently in execution on a given core. Thanks to this priority, the run-time is capable of allocating more power to the higher priority process~\cite{bartolini2017self,CESARINI_VLSISOC}. In ANTAREX, these APIs can be inserted by LARA aspects in the application code.




\section{Evaluation}
\label{sec:eval}
Tables~\ref{tab:staticMetrics} and~\ref{tab:dynamicMetrics} show static and dynamic metrics collected for the weaving process of the presented examples. 
In Table~\ref{tab:staticMetrics}, we can see the number of logical lines of source code for the LARA strategies, as well as for the input code and generated output code (the SLoC-L columns). In the last two columns we report the difference in SLoC and functions between the input and output code (the delta columns). Note the woven and delta results for the HalfPrecisionOpenCL strategy are the sum of all generated code, totaling 31 versions.

An inspection of columns LARA SLoC-L and Delta SLoC-L reveals that, in most examples, there is a large overhead in terms of LARA SLoC-L over application SLoC-L. 
While this may seem a problem, we need to consider that a large part of the work being performed by these strategies is code analysis, which does not translate directly to SLoC-L in the final application. 
Furthermore, the Delta SLoC-L metric does not account for removed application code and for these cases a metric based on the similarity degree among code versions could be of more interest. 
Also, in real-world applications, the ratio of LARA SLoC-L to application SLoC-L would be definitely more favorable, thanks to aspect reuse.

\begin{table*}[htb!]
\scriptsize
\centering
\caption{Static Metrics}
\label{tab:staticMetrics}
\begin{tabular}{lllllllll}
Strategy             & LARA & LARA & Input & Input & Woven & Woven & Delta & Delta \\
 & SLoC-L & Aspects & SLoC-L & Func & SLoC-L & Func & SLoC-L & Func \\
ChangePrecision      & 27          & 1            & 12           & 3               & 13           & 3               & 1            & 0               \\
SimpleExamon         & 20          & 1            & 12           & 3               & 23           & 5               & 11           & 2               \\
Multiversion         & 46          & 2            & 12           & 3               & 43           & 5               & 31           & 2               \\
CreateFloatVersion   & 28          & 2            & 12           & 3               & 24           & 3               & 12           & 0               \\
SimpleLibVC          & 12          & 1            & 12           & 3               & 39           & 4               & 27           & 1               \\
HalfPrecisionOpenCL* & 93          & 3            & 9            & 1               & 279          & 31              & 270          & 30              \\
Total                & 226         & 10           & 69           & 16              & 421          & 51              & 352          & 35             
\end{tabular}
\end{table*}

\begin{table}[htb!]
\scriptsize
\centering
\caption{Dynamic Metrics}
\label{tab:dynamicMetrics}
\resizebox{1\columnwidth}{!}{
\begin{tabular}{llllll}
File                & Selects & Attributes & Actions & Inserts & Native \\
  & & & & & SLoC\\
ChangePrecision     & 4       & 109        & 2       & 1       & 0           \\
SimpleExamon        & 4       & 131        & 18      & 7       & 0           \\
Multiversion        & 8       & 477        & 27      & 16      & 9           \\
HalfPrecisionOpenCL & 125     & 2211       & 381     & 159     & 31          \\
CreateFloatVersion  & 2       & 170        & 6       & 3       & 0           \\
SimpleLibVC         & 7       & 93         & 13      & 8       & 36          \\
Total               & 150     & 3191       & 447     & 194     & 76         
\end{tabular}
}
\end{table}

To better understand the impact of analysis, we report in the first two columns of Table~\ref{tab:dynamicMetrics} the number of code points and of their attributes analysed, which can be compared with the last three column of the same table, which instead report the corresponding effects, in terms of the number of modified points and lines of code inserted.
To understand the impact of removed lines of code, we look at the \emph{Inserts} and \emph{Actions} columns, which show that circa one half of the actions do not insert code.
The end line is that the analysis work exceeds the transformation work by an order of magnitude, and the insertions only underestimate significantly the work performed.

Another benefit for user productivity when using LARA is how the techniques presented in the examples can scale into large-scale applications and scenarios. Most of the presented strategies are parameterized by function, i.e., they receive a function join point or name and act on the corresponding function. This could be performed manually, albeit crudely, using a search function of an IDE. Consider the case where we instead want to target a set of functions, whose names we may not know, based on their function signatures, or based on the characteristics of the variables declared inside their scope. This kind of search and filtering based on syntactic and semantic information available in the program is one of the key features of LARA and it cannot be easily attained with other tools. As the aspects presented here illustrate, LARA strategies can be made reusable and applied over large applications, greatly out scaling the effort needed to develop them.

\section{Case Study: Computer Accelerated Drug Discovery}
\label{sec:appl1}

In this section, we provide an overview of the application of the ANTAREX tool flow to the Drug Discovery application developed in Use Case 1.

Computational discovery of new drugs is a compute-intensive task that is critical to explore the huge space of chemicals with potential applicability as pharmaceutical drugs.
Typical problems include the prediction of properties of protein-ligand complexes (such as docking and affinity) and the verification of synthetic feasibility.

In the ANTAREX project it is expected that this application supports the exploration of millions of ligands over several pockets with sizes ranging from a few thousand atoms to several tens of thousands, in a timely manner.
Performance is a key factor for this application, and we have used the ANTAREX tool flow to optimize the use case for Exascale systems.

\subsection{Auto-parallelization and exploration through the ANTAREX DSL}

Experiments on using an entire cluster for execution of the Use Case 1 have shown that relying only on MPI to parallelize the code both to distribute parallelism between nodes and to take advantage of parallelism inside a node is highly inadequate. 
We observed that a few tens of thousand tasks are enough to make MPI fail during execution. To overcome that, we opted to use MPI only for distributing work between nodes, and parallelize the application inside the node with OpenMP.

We have developed an auto-parallelization LARA library that uses static analysis to detect and insert OpenMP pragmas in \texttt{for} loops that can be parallelized.
We have identified the function \texttt{MeasureOverlap} in the UC1 application as an important hotspot, and we have extracted it to a test application written in C (the auto-parallelization library currently only supports C).

\Cref{fig:par-outer-loops} shows the LARA code that parallelizes the application. Line 6 calls the parallelization library, which parallelizes all the \texttt{for} loops in the code that where detected as safe to be parallelized.
However, this can be a very inefficient strategy, since nesting parallelism with OpenMP can lead to excessive overhead.

The code in lines 9-28 detects and disables all nested OpenMP pragmas. It starts by iterating over all OpenMP pragmas in the code of the kind \texttt{parallel for}; then, it retrieves all descendants of the loop corresponding to that pragma, that are also OpenMP pragmas of the kind \texttt{parallel for}, and transforms those nested pragmas into inlined comments.

\begin{figure}[ht]
	\begin{center}
\begin{lstlisting}[boxpos=b,style=LARA]
import clava.autopar.Parallelize;

aspectdef ParallelizeOuterPragmas

	// Parallelize all loops
	Parallelize.forLoops();
	
	// Disable nested OpenMP pragmas
	select omp end
	apply
		if($omp.kind !== 'parallel for') {
			continue;
		}
		
		if($omp.target === undefined) {
			continue;
		}
		
		// 'target' is the statement associated with the pragma
		// (a 'for' loop in this case)
		for($innerOmp of $omp.target.descendants('omp')) {
			if($innerOmp.kind !== 'parallel for') {
				continue;
			}
			// Transform the inner OpenMP pragma into an inlined comment
			$innerOmp.replaceWith('// ' + $innerOmp.code);
		}
	end
end
\end{lstlisting}
		\caption{LARA code for loop parallelization.}
		\label{fig:par-outer-loops}
	\end{center}
\end{figure}

After parallelizing the code, we want to explore two parameters: the number of threads to use, and the size of the pocket to test.
\Cref{fig:lara-uc1-dse} shows the LARA code for a strategy that uses the LARA library LAT\footnote{LAT is available at \href{https://github.com/specs-feup/LAT-Lara-Autotuning-Tool}{https://github.com/specs-feup/LAT-Lara-Autotuning-Tool}} to explore these parameters.
The strategy sets the number of threads and the number of atoms of the pocket size to explore, and creates versions of the code for all combinations of these parameters.
For each combination, the strategy will compile, run and collect the results, which are aggregated into a CSV file at the end.

In more detail, line 10 creates an instance of the Lat class, which will be used throughout the strategy.
Since the code has been instrumented with OpenMP pragmas, line 13 sets some necessary compilation flags.
Line 16 select the section of code we want to explore (i.e., the call to the function \texttt{MeasureOverlap}), and lines 17-20 set the call as the place in the code where we want to measure the metrics, and the body of the function where the call appears as the place where the parameters will be changed.

Lines 26-28 create the ranges of values we want to explore: a set of predefined values for the variable \texttt{num\_pocket\_atoms}, and a set of number of threads that start at 1 and double at each step.

Lines 33-34 set the metrics we want to measure (i.e., execution time and energy), line 37 sets the number of times we should repeat the measures for each version of the code, and line 38 executes the exploration.
Finally, line 39 collects the results to a CSV file.


\begin{figure}[ht]
	\begin{center}
\begin{lstlisting}[boxpos=b,style=LARA]
import lat.Lat;
import lat.LatUtils;
import lat.vars.LatVarOmpThreads;

import lara.metrics.EnergyMetric;
import lara.metrics.ExecutionTimeMetric;

aspectdef UC1Exploration
	
	var lat = new Lat('uc1_exploration');
	
	// Compiler flags
	lat.loadCmaker().cmaker.addFlags('-fopenmp', '-O3', '-march=native');

	// Scope for variable change
	select function{"main"}.body.call{"MeasureOverlap"} end
	apply
		lat.setMeasure($call);
		lat.setScope($body);		
	end
	
	// Design space
	pocketSizes = [5000, 7000, 10000, 12000, 50000];
	maxThreads = 5;

	var samples = new LatVarList("num_pocket_atoms", pocketSizes);
	var threadValues = new LatVarRange("numThreads", 0, maxThreads, 1, 
                                     function(x){return Math.pow(2, x);});
	var numThreads = new LatVarOmpThreads(threadValues);
	
	lat.addSearchGroup([samples, numThreads]);

	// Metrics
	lat.addMetric(new ExecutionTimeMetric());
	lat.addMetric(new EnergyMetric());

	// Tune
	lat.setNumTests(5);
	var results = lat.tune();
	LatUtils.resultToCsv(results, './');
end
\end{lstlisting}
		\caption{LARA code for parameter exploration.}
		\label{fig:lara-uc1-dse}
	\end{center}
\end{figure}




\subsection{Performance}


We have applied the auto-parallelization and exploration strategy to the C version of the \texttt{MeasureOverlap} function, considering pocket sizes from 5.000 atoms to 50.000 atoms.
We run the experiments on a computer with an Intel Xeon CPU E5-1650 v4 processor, at 3.60GHz and 64GB of RAM. The code was compiled with gcc 7.3.0.

\begin{figure}[ht]
	\centering
	\resizebox{0.9\columnwidth}{!}{\includegraphics{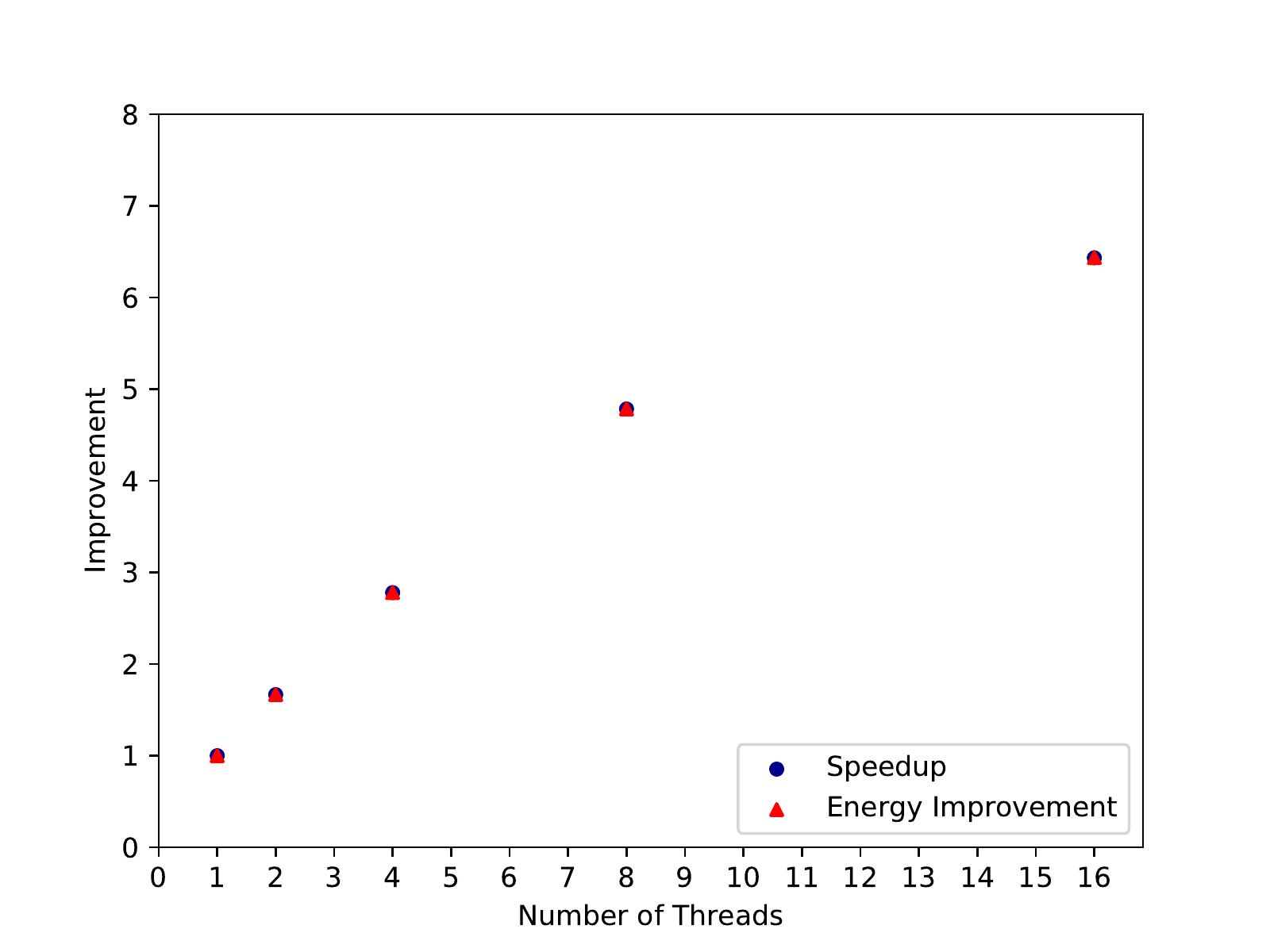}}
	\caption{\label{fig:uc1-results}Results of the DSL-enabled automatic exploration}
\end{figure}

\Cref{fig:uc1-results} shows the improvements in execution time (i.e., speedup) and energy (i.e., energy improvement) when the number of threads changes, when considering the biggest pocket size (15.000 atoms). We use as input the sizes of 120k ligands, based on a database provided by the industrial partner.

As expected for this case, the improvement curve of the execution time coincides with the improvement in energy consumed, and increases with the number of threads.
The number of atoms in the pocket size modestly changes the speedup, i.e. between 0.7 \% and 5.5 \%, for the same number of threads.
The amount of change increases with the number of threads, but for a given number of threads, an increase in pocket size improves speedup up to a point, and after that the speedup decreases (we consider that this happens due to saturation of memory resources).
This kind of exploration can be used to generate data which can be fed to the autotuner.

\section{Case Study: Self-Adaptive Navigation System}
\label{sec:appl}
In this section, we provide an overview of the application of the ANTAREX tool flow to the Self-Adaptive Navigation System developed in Use Case 2.
The system is designed to process large volumes of data for the global view computation and to handle dynamic loads represented by incoming routing requests from users of the system. Both disciplines require HPC infrastructure in order to operate efficiently while maintaining contracted SLA. Integration of the ANTAREX self-adaptive holistic approach can help the system to meet the mentioned requirements and pave the way to scaling its operation to future Exascale systems.

Core of the system is a routing pipeline with several stages which uses our custom algorithm library written in C++. The library provides an API for the individual routing algorithms and for data access layer, which provides abstraction of a graph representation of the road network. The graph is stored in a HDF5 file, which is a well known and convenient storage format for structured data on HPC clusters.

\subsection{Memoization and Precision Tuning through the ANTAREX DSL}
The navigation system in our use case calculates routes simultaneously for possibly all participants in the system and it much relies on precise calculation of the macroscopic traffic flow model. The critical component for our traffic flow calculation is the \emph{Betweenness Centrality} algorithm, which also poses the biggest challenge for computation efficiency.
The Betweenness Centrality algorithm can alternatively be used as a tool for a what-if analysis on a real traffic network~\cite{hanzelka2018betweenness}. 

We have applied two source-to-source transformations to the Betweenness Centrality algorithm: custom precision and memoization. 
The original application contains 770 lines of the code (C code + the header files). 
The LARA aspect in Figure~\ref{fig:lara:memoize} shows the main strategy for the custom precision transformation. 
It replaces the references to \texttt{double} type by a new symbol and replaces all the syntactic expressions that depend on the \texttt{double} type (for instance, mathematical functions) by new symbols. 
The custom precision library generates several files that allow to have different definitions for the created symbols: a file that defines the created parameter symbols by the original values (\texttt{double} here); a file that defines the created symbols to have a \texttt{float} version, generated by the call to the \texttt{CustomPrecisionGenParametersFor('float')} aspect; a file that contains a template to have another type representation. 
The last one must be updated by the user to a specific type representation (for example: half precision, fixed point representation, etc.). 
After applying the custom precision strategy, the final number of lines of the Betweenness Centrality application is equal to 870.

\begin{figure}[ht]
	\begin{center}
\begin{lstlisting}[boxpos=b,style=LARA]
import clava.Clava;
import clava.ClavaJoinPoints;
import antarex.utils.messages;
import antarex.precision.CustomPrecision;

aspectdef Launcher
	println('ANTAREX UC2, Custom precision')
	call CustomPrecision_Initialize();
	call CustomPrecisionFor('double');
	call CustomPrecisionGenParametersFor('float');
	call CustomPrecision_Finalize();
end

\end{lstlisting}
	\end{center}
\caption{Custom precision LARA launcher for Betweenness application}
\label{fig:lara:memoize}
\end{figure}

For the memoization strategy, which was used as a proof-of-concept for this technology, the main DSL aspect is shown in Figure~\ref{fig:lara:memo2}. 
It applies the memoization to the method \texttt{computeMetric} of the \texttt{Betweenness} class.

\begin{figure}[ht]
	\begin{center}
\begin{lstlisting}[boxpos=b,style=LARA]
import clava.Clava;
import clava.ClavaJoinPoints;
import antarex.utils.messages;
import antarex.memoi.Memoization;

aspectdef Launcher
	println('ANTAREX UC2, Memoization')
	call Memoize_Initialize();
	call Memoize_Method('Betweenness', 'computeMetric');
	call Memoize_Finalize();
end

\end{lstlisting}
	\end{center}
\caption{Lara launcher for memoization in Betweenness application}
\label{fig:lara:memo2}
\end{figure}

After applying the strategy, the application has 775 lines of code, and uses a custom memoization library automatically generated by Clava with around 150 lines of C code. The number of generated lines depends on the number of memoized functions (one in this case). 
So, the total number of lines is 925, considering the memoization library.
In order to optimize the memory allocation, we applied another modification to the Betweenness Centrality algorithm. It consists of moving statements outside the body of a loop without affecting the functionality of the program (also called code hoisting). 
This transformation was applied manually and is not defined in LARA.
We also applied the memoization technique to the routing algorithms for one function, with a similar number of generated lines as the Betweenness Centrality algorithm.

\subsubsection{Performance and scalability}
For testing impact of INRIA tools on Betweenness Centrality algorithm, we chose a graph of the traffic network of \emph{Porto} city in Portugal and one bigger graph of 4 countries (\emph{4EU} - Czech Republic, Slovakia, Austria, Hungary). 
This experiment was focused on comparing the speed of an original and of a modified version of the Betweenness Centrality algorithm. 
The size of both graphs in terms of vertex and edge count is presented in Table~\ref{tab:memo:res}.

\begin{table}
\label{tab:memo:res}
\centering
\caption{Characterization of the traffic network graphs used in the experiments}
\begin{tabular}{crr} 
Graph & Vertices & Edges\\
\emph{Porto} city area &	7,316 & 	14,793\\
\emph{4EU} countries & 	3,567,735 &	8,610,752\\
\end{tabular}
\end{table}

The performance was tested on Salomon cluster operated by the IT4Innovations National Supercomputing Center. 
The cluster consists of 1,008 compute nodes and each of them contains 2x Intel Xeon E5-2680v3 processors clocked at 2.5 GHz and 128 GB DDR4@2133 RAM. When the experiments were performed, the operating system was CentOS 7.6 and the code was compiled using Intel C++ Compiler 17.0.

We tested several versions of the Betweenness Centrality algorithm with various combinations of the ANTAREX tools:
\begin{description}
\item[Float (F)] a version of the algortihm with reduced precision of internal variables
\item[Double (D)] a standard algorithm
\item[Float\_Hoisting (FH)] a float version with hoisting
\item[Double\_Hoisting (DH)] a standard version with hoisting
\item[Float\_Hoisting\_Memoization (FHM)] a float version with hoisting and memoization
\item[Double\_Hoisting\_Memoization (DHM)] a standard version with hoisting and memoization
\end{description}

Table~\ref{tab:betwc:porto} shows run times with the optimizations provided by different ANTAREX tools for different number of nodes. 

\begin{table}
\centering
\label{tab:betwc:porto}
\caption{Betweenness Centrality algorithm run times for the Porto graph (in seconds)}
\begin{tabular}{crrrrrr}
Compute Nodes & F & FH & FHM & D & DH & DHM \\
1 & 1.20 & 1.15 & \emph{1.12} &1.34 &1.27 &1.25 \\
2 &0.66 &0.65 &\emph{0.64} &0.75 &0.72 &0.71 \\
4 &\emph{0.41} &0.42 &\emph{0.41} &0.47 &0.45 &0.44 \\
\end{tabular}
\end{table}

\begin{table}
\caption{Betweenness Centrality algorithm run times for the 4EU graph (in $10^3$ seconds)}
\begin{tabular}{rrrrrrr}
Compute Nodes  & F     & FH           & FHM        & D          & DH         & DHM         \\
1     & 294.575   & 283.534 & \emph{282.233} & 340.523 & 325.428 & 315.872 \\
2     & 153.296   & 150.123 & \emph{147.966} & 175.940 & 170.587 & 167.524 \\
4     & 79.484    & 79.233  & \emph{75.314}  & 89.905  & 87.206  & 86.106  \\
8     & 40.811    & 39.899  & \emph{38.711}  & 46.031  & 44.736  & 44.258  \\
16    & 21.305    & 21.070  & \emph{20.247}  & 23.725  & 22.958  & 22.493  \\
32    & 11.122    & 10.809  & \emph{10.590}  & 12.228  & 11.884  & 11.791  \\
64    & 5.922     & 5.726   & \emph{5.677}   & 6.489   & 6.284   & 6.092   
\end{tabular}
\end{table}

Overall, we can see a speedup of 3.7 to 7.8\% for the larger graph when adding the hoisting and memoization transformations to the double precision floating point code.
Considering also the precision reduction, the performance speedup grows to 14.3 to 20.6\% depending on the number of cores employed.

\subsection{Autotuning for Client-side navigation}
The Navigation application is the one which finally sells the whole navigation system, so it requires quite an attention. In a holistic approach the application implementation needs to consider multiple aspects, both in the functional and the performance space as was stipulated in the specification of the system.
In simple words: 
\begin{itemize}
\item Navigation shall not exceed defined data consumption
\item Navigation shall contribute with car data to build the knowledge of the server-side system
\item Navigation secures the required navigation quality
\item Navigation optimizes communication with server-system to alleviate it from over loading
\end{itemize}
All these goals have been addressed through the integration of a run-time autotuner.

Important from the HPC perspective is that a navigation client acts as an intelligent source of computational demand, able to regulate the number of requests based on the quality of the navigation and the ability to perform local route computation. 

\begin{figure}[ht]
\centering
\includegraphics[width=1\textwidth]{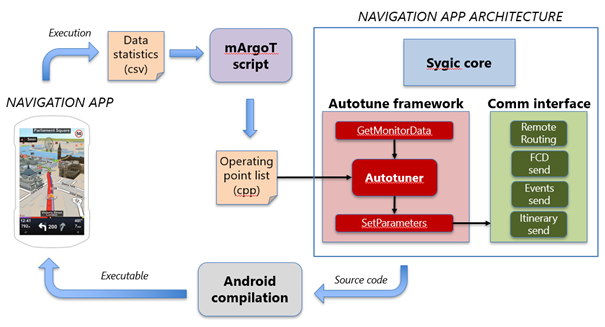}
\label{fig:uc2:flow}
\caption{Autotuning framework application to the navigation app}
\end{figure}

Figure~\ref{fig:uc2:flow} shows the tool-flow for configuring the autotuner towards statistically the best performance. The input is the data consumption statistics on estimated data transfer per request type for all service modules, while collected from prior test runs (circa 100 hours) of the navigation app under prevailing commute conditions (in Bratislava city). 
As the autotuner architecture evolved the definition of the input has been, in addition to mean, extended to contain the mean, max, min, and the standard deviation of the variables.

\begin{figure}[ht]
\centering
\includegraphics[width=0.8\textwidth]{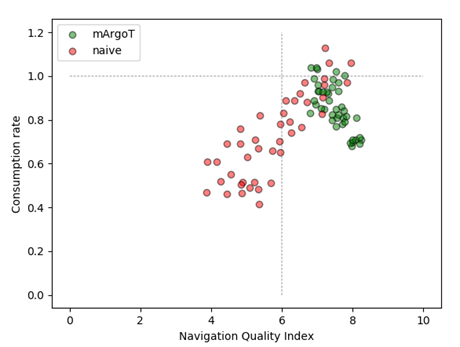}
\label{fig:uc2:perf}
\caption{Performance comparison between baseline and mArgot autotuning}
\end{figure}

Figure~\ref{fig:uc2:perf} shows the effectiveness of the mArgot autotuner compared with the baseline autotuner included in the commercial release of Sygic Truck Navigation v13.7.0 (March 2017).
The baseline (red dots) is a simple autotuning implementation that supports only a data transfer limit set at 20MB.
It does not support constraints on the navigation quality, providing the maximum quality based on its prediction of data usage.

The mArgot autotuner (green dots) instead takes into account a \emph{Navigation Quality Index} (NQI) as a constraint.
The NQI is defined as a fuction of the remote routing frequency, which saturates at a point dependent on the current traffic level.
The constraint in the proposed experiment is for the NQI to stay above a minimum of 6.0.
Furthermore, it uses the same data transfer limit (20MB), a minimal data contribution of 20\%, and an estimate of 40 monthly driving hours.

As shown in Figure~\ref{fig:uc2:perf} the objectives are well met with an exception of few dots about ideal consumption rate, which can be explained with stochasticity of driver behaviour records (data generated only in the case of excessive acceleration, braking, cornering).

\begin{figure}[htb]
\centering
\includegraphics[width=0.8\textwidth]{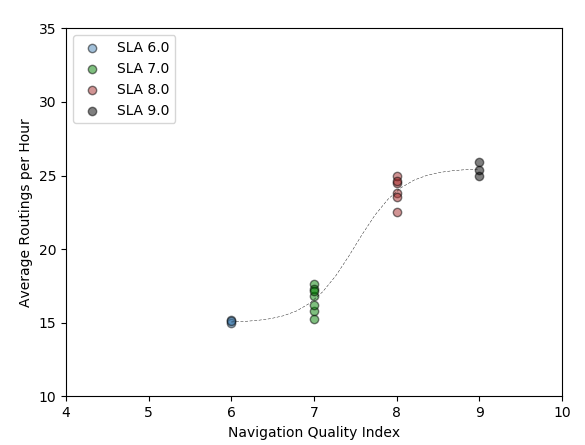}
\label{fig:uc2:sla}
\caption{Variation of the compute workload (in routing requests per hour) as a function of the Navigation Quality Index for a scenario of medium traffic commute in Bratislava}
\end{figure}
The impact of the NQI constraint selection can be understood through the experiments reported in Figure~\ref{fig:uc2:sla}.
In the experiment, the NQI is set to different values in the range [6.0,9.0]. 
The results show how relaxing the quality level allows to save computational resources, lowering the total cost of routing, e.g., by 12\% when dropping from NQI 8.0 to 7.0.  
The results also allow to identify the best trade-off from a value-at-cost solution in practical cases, at NQI 6.8.
Compared to the baseline autotuning, which achieves an NQI of 6.25, the mArgot autotuner allows a saving of 14\% in computational resources while providing a better quality of service, assuming the use of the best trade-off NQI of 6.8.

\section{Related Works}
\label{sec:relw}
Many Domain Specific Languages have been proposed for the domain of High Performance computing~\cite{giles2014trends}. 
A survey of tools and approach in generative programming for high performance computing can be found in~\cite{cohen2006search}, whereas a collection of techniques for and examples of modern DSL design can be found in~\cite{mernik2012formal}.

Many of them actually focus on a specific class of HPC applications, either in terms of a specific field such as computational biology~\cite{starruss2014morpheus}, or a specific mathematical problem, such as partial differential equations~\cite{devito2011liszt,rathgeber2012pyop2,schmitt2014exaslang,membarth2012towards}, often using stencil computation, i.e., common functionalities in a given field implemented in an optimised way and offered through the DSL to the user, possibly with some autotuning capabilities.
This approach is effective in terms of combining efficiency of the generated code with familiarity of the language for the domain expert.
However, a different solution is needed for each domain or problem, and when HPC experts at a supercomputing center are called for to optimise a customer's code, this forces specialisation, which may be undesirable from the point of view of the supercomputing center management. 

Indeed, at least some such centers have attempted to develop a coherent set of languages and tools applicable across multiple application domain within the wider context of HPC~\cite{duran2011ompss}, and large scale research programs have been devoted to this goal\cite{weiland2007chapel}.
These DSLs are also related to algorithmic skeleton frameworks~\cite{gonzalez2010survey}, which however do not take into account the presence of the HPC expert.
Model-driven approach have been also proposed~\cite{palyart2011mde4hpc,palyart2012hpcml}, with the goal to abstract the application from the concrete and ever-changing HPC infrastructure.
A more general-purpose take on the stencil computation is Terra~\cite{devito2013terra}, which employs Lua to support the creation of DSL compilers which can interoperate with the generated code.

All these efforts have the opposite problem -- they force the application developer to acquire skills typical of the HPC experts.
Furthermore, they do not deal easily with the large amount of legacy code found in HPC, where applications may remain in use for decades~\cite{giles2014trends,palyart2012hpcml}.
The ANTAREX DSL, on the other hand, is designed exactly to cater to these needs.
By creating an external set of LARA aspects, the ANTAREX DSL strategies, the HPC expert can manipulate the code developed by the application domain expert. 
The DSL code can be as small or as large as needed to reach the performance goals, but never intrudes in the ability of the application developers to maintain or extend their applications. 
At the same time, by providing access to commonly used computation patterns, code transformations, and autotuning tools, the ANTAREX DSL enables the HPC experts to reuse solutions across different applications, removing the need to reinvent the wheel with every new application domain.

\section{Conclusions}
\label{sec:conc}
%
%

In this paper, we have reported the major concrete outcome of the ANTAREX research project, the ANTAREX DSL.
The DSL, based on the AOP language LARA, aims at supporting a well-established collaboration model between users and support staff of HPC centers, where the users develop applications focusing on the functional requirements, and then take advantage of the support staff's expertise on HPC to improve the performance of the applications. 
With the ANTAREX DSL, this process is simplified by removing the need for the HPC support staff to manually modify the user's code, instead relying on aspects which encode available code transformations as well as providing access to libraries for autotuning, system monitoring, and dynamic recompilation.

We demonstrate the use of our tools in a scenario where the HPC center runs remote navigation for a pool of users, considering both urban and transnational scenarios, as well as in a scenario where the HPC center runs a drug discovery appplication.

\section*{Acknowledgements}
The ANTAREX project is supported by the EU H2020 FET-HPC program under grant 671623.
The IT4Innovations infrastructure is supported by the Large Infrastructures for Research, Experimental Development and Innovations project ``IT4Innovations National Supercomputing Center -- LM2015070''. 
The CINECA infrastructure is supported by the EU and the Italian Ministry for Education, University and Research (MIUR) under the PRACE project.

\bibliographystyle{IEEEtran}
{\footnotesize
\bibliography{biblio,lara,memoi}
}

\end{document}